\documentclass[12pt,a4paper]{article}
\usepackage[dvips]{graphicx}
\usepackage{rotate}
\usepackage{epsf}
\usepackage{cite}
\usepackage{latexsym}
\usepackage{amsmath}
\usepackage{lscape}
\usepackage{rotating}
\usepackage{amssymb}
\usepackage{amsfonts}

\newcommand{\be}{\begin{equation}}
\newcommand{\ee}{\end{equation}}
\newcommand{\bea}{\begin{eqnarray}}
\newcommand{\eea}{\end{eqnarray}}
\newcommand{\eg}{ \end{gather} }
\newcommand{\bg}{\begin{gather}} 
\newcommand{\nn}{\nonumber}

\newcommand{\obrx}[4]{\begin{figure}[htbp] \centering \begin{minipage}{0.8\textwidth} \centering \scalebox{#1}{\includegraphics{#2}} \caption{\normalsize \em #3} \label{#4} \end{minipage} \end{figure}}
\newcommand{\ud}{\mathrm{d}} % v integralu
\newcommand{\ue}{\mathrm{e}} % exponenciela
\newcommand{\celE}[1]{\mathrm{E}\! \left(#1 \right)} % elipticky integral
\newcommand{\celK}[1]{\mathrm{K}\! \left(#1 \right)} %elipticky integral
\newcommand{\dd}[2]{\frac{\ud #1}{\ud #2 }} % derivace #1 podle #2  d#1/d#2
\newcommand{\ddp}[2]{\frac{\partial #1}{\partial #2 }} % parc der #1 podle #2
\newcommand{\pp}{\frac{\pi}{2} }
\def\p{\pi}
\def\pr{^{\prime}}
\newcommand{\cl}[1]{\ensuremath{ \mathcal{#1}}}
\newcommand{\Ce}{\ensuremath{ \mathcal{C} \,}} % krivka stability
\newcommand{\ada}{\frac{\phi_D}{\phi}}
\newcommand{\nmne}{\ensuremath{ \frac{n_e}{n_m}}}
\newcommand{\re}{\ensuremath{\mathcal{R}e \,}}
\newcommand{\im}{\ensuremath{\mathcal{I}m \,}}

\newcommand{\abs}[1]{\left| #1\right|}

\def\half{\frac12}
\def\eip{\ue^{i  \vf}}
\def\a{\alpha}

\def\d{\delta}
\def\ve{\varepsilon}
\def\f{\phi} % \varphi
\def\vf{\varphi}

\def\k{\kappa} % Also, \varkappa (see below)

\def\p{\pi}     % Also, \varpi
\def\t{\tau}

\def\O{\Omega}
\def\P{\Pi}

\def\S{\Sigma}

\def\cf{\mathcal{F}}
\def\ce{\mathcal{E}}
\def\ch{\mathcal{H}}

\begin{document}
\begin{titlepage}

\begin{center}
\vskip3em
{\large\bf Towards the Core of the\\
\vskip .5em
Quantum Monopole}

\vskip3em
{A.~Campbellov\'a,\footnote{E-mail:campbell@physics.muni.cz}\\
\vskip .5em
R.~von~Unge,\footnote{E-mail:unge@physics.muni.cz}}\\
\vskip1em {\it Institute for Theoretical Physics and Astrophysics\\
Masaryk University\\
Kotl\'a\v{r}sk\'a 2\\
611 37 Brno\\
Czech Republic}
\end{center}

\vfill

\begin{abstract}

\noindent
We study monopole solutions of the quantum exact low-energy effective  $N=2$
super Yang-Mills theories of Seiberg and Witten. We find a first order differential
equation for the spatial dependence of the moduli and show that it can be
interpreted as an attractor equation. Numerically integrating this equation, we try to
address the question of what happens when one approaches the quantum core of
the monopole where the low energy effective theory breaks down or,
alternatively, if there are modified monopole solutions that do not have a
strongly coupled quantum core so that one may trust the solution not only
asymptotically.
\end{abstract}

\vfill
\end{titlepage}
\section{Introduction}
In the paper \cite{Chalmers}, quantum corrected BPS monopole solutions in ${\cal
N}=2$ supersymmetric Yang-Mills theory were found. Using the exact low energy
effective Hamiltonian derived from the Seiberg-Witten solution of the effective action
\cite{SeiWitt} it was shown that as one moves towards the center
of the monopole, the moduli of the solution change in exactly such a way as
to keep the phase of the local central charge constant. As was shown in \cite{SooJong,Denef1,Denef2},
this result can be thought of as a rigid,
nongravitational counterpart of the attractor mechanism  \cite{Attractor1,Attractor2} and as such
it serves as a toy model for the full gravitational attractor mechanism.

However, while the original attractor mechanism is purely classical and perturbative
quantum corrections have only recently been considered
\cite{Gunaydin1,Fandoren,Alexandrov,Gunaydin2},
in the gauge theory case the theory is fully quantum, both perturbative and nonperturbative.
One might therefore hope to learn more about quantum corrections to the 
attractor mechanism by studying this toy model.

There is also another motivation for further studying the description of quantum corrected
monopoles given in \cite{Chalmers}.
As one approaches the center of the monopole the moduli reaches the strong
coupling region where the effective action description ceases to be valid
and one cannot trust the solution anymore. In this paper we investigate
possible ways out of this dilemma. One possibility would be to mimic the
enhancon idea \cite{Johnson} so that one cuts off the solution at some finite
radius and replaces the center with another weakly coupled solution.
Another possible solution would be to use a duality transformation to
change the strongly coupled description in the core to a weakly coupled
dual description.

Indeed in this paper we argue that a scenario somewhat analogous to the enhancon mechanism
is possible to realize.
By appropriately choosing an integration constant of
the BPS equation (which is forced to be zero in the classical 't Hooft Polyakov solution),
we are able to find a natural cutoff point in the weak coupling region. At the cutoff radius
both electric and magnetic fields are zero and there is no contribution to the the mass of
the monopole from the inner boundary. The center gets replaced by a bubble of
Higgs vacuum. The similarity to the enhancon mechanism is only partial however, in our
scenario we see no sign of symmetry enhancement. In fact, while the abelian parts of the gauge
field goes to zero at the cut off radius, the nonabelian parts (W-bosons) stay massive and
nonzero. Also, one might have hoped that the quantum corrections would have made the
geometry completely nonsingular (along the lines of \cite{nonsingmpole}) but the energy
density will be discontinuous indicating the presence of a shell like singularity.

This paper is organized as follows: in section 2 we review the basic results of
\cite{Chalmers}. In section 3 we derive a differential equation for the spatial dependence
of the moduli of the solution and show how this equation is related to the attractor mechanism.
In sections 4 and 5 we discuss the general behavior of the various fields as well as the energy
density when one moves towards the center of the monopole. In section 6 we then
give numerical results for various special cases and discuss the various scenarios
that appear and the possibility to enhance them to solutions valid everywhere. Finally we
conclude in section 7. In an appendix various useful expansions around the strong coupling singularity
(the attractor point) $u=1$ are discussed.

\section{Review}
The leading term of the low energy effective action of $\cl{N}=2$
super Yang-Mills theory is determined by a holomorphic function $\cal{F}(W)$ of the $\cl{N}=2$ gauge superfield $W$ 
\be
\cl{S}_\cl{F}=\frac1{2\p} \im \int \ud ^4 x \, \ud^4 \theta \, \cl{F}(W).
\ee
Reducing the \cl{N}=2 action to \cl{N}=1 formulation and then to \cl{N}=0 language we find for the bosonic part the action
\begin{gather*}
\cl{S}_{\cl{F}} \!= \!-\frac1{4\p}  \, \im \!\! \int \!\! \ud^4 x  \,
\cl{F}_{AB}\, \left[ \frac12 (B_i^A+iE_i^A)(B_i^B+iE_i^B)+\nabla_{\mu} \phi^A\nabla^{\mu} \bar \phi^B +  \right.\\
\left. +\frac12 [\phi,\bar \phi]^A [\phi, \bar \phi]^B \right],
\label{bosonicaction}
\end{gather*}
where the magnetic and electric field are components of the field strength tensor $B_i^A=\epsilon_{ijk} F^{jk\,A}$ and $E_i^A=F_{0i}^A$, $\f$ is the complex scalar, 
 $A=1,2,3$ is the $SU(2)$ group index and $\cl{F}_A$ and $\cl{F}_{AB}$ are the first and second derivative of the prepotential 
$\cl{F}_A=\ddp{\cl{F}}{\phi^A}$, $\cl{F}_{AB}(\phi)=\ddp{\cl{F}}{\phi^A \partial \phi^B}$.
We consider only static configurations and choose the gauge $\nabla_0 \phi^A=0$. \\

In order to have finite energy configurations the vacuum expectation value for $r \rightarrow \infty$ must approach the Higgs vacuum, i.e. the potential must be zero and the scalar field must commute with its conjugate $[\phi, \bar \phi]=0$. We can write the expectation value using a unit vector in the form 
$$
<\phi ^A> = a \ue ^A.
$$
Then since \cl{F} must be gauge invariant, it can depend only on $\phi^2=\S_{A=1}^3 \phi^A \phi_A$ and the derivative $\cl{F}_A$ is simply
\be
\cl{F}_A=\ddp{\cl{F(\phi)}}{\phi} \frac{\phi^A}{\phi} =\cl{F}' \frac{\phi^A}{\phi}.
\ee 
The expectation value of the dual field $\phi_{D\,A}=\cl{F}_A$ is
$$
<\phi_{D \, A}> = \cl{F}' \ue_A \equiv a_D \, \ue_A.
$$
The parameter $a$ is not gauge invariant, under the Weyl group it changes its sign $ a \leftrightarrow -a$. Classically, a suitable gauge invariant parameter to distinguish the various Higgs vacua is $u=<\frac12 \phi^2>$.\\

It was shown in\cite{Chalmers} that as long as the imaginary part of $\cl{F}_{AB}$ is positive
it is possible to write the effective the Hamiltonian as a positive definite part plus
a total derivative. Moreover, putting the positive definite part to zero one gets a
BPS equation for the general monopole which is the same as for the classical theory
\be
B_j^A+iE_j^A+\ue^{i \alpha} \sqrt{2} \nabla_j \phi^A =0, \label{BPS}
\ee
where $\ue^{i \alpha}$ is a constant phase. When the BPS equation is satisfied the whole
contribution to the energy is from the total derivative term and solutions satisfying this
equation are called BPS solutions.
We define the electric and magnetic quantum numbers by integrals at spatial infinity as
\begin{align}
n_e a= -\int \ud \vec{\Sigma} \cdot \vec{\Pi}_A \phi^A,&
& n_m a_D=-\frac1{4\p} \int \ud \vec{\Sigma} \cdot \vec{B}^A \phi^D_A,
\end{align}
where $\vec{\Pi}_A$ is the momentum conjugate to the gauge field $\vec{A}^A$
\be
\vec{\Pi}_A=-\frac1{4\p} \re \left\{ \cl{F}_{AB} \left(\vec{B}^B+i\vec{E}^B\right) \right\}.
\label{Pi}
\ee
The contribution to the mass from the total derivative term can be written as
\be
\im e^{i\alpha}Z,
\ee
for $Z=n_e a+n_m a_D$. This contribution to the energy will have its maximum value
if we choose the phase $\ue^{i \alpha}$ to be related to the central charge $Z$ by
\be
\ue^{i \alpha} = i \frac{ \bar{Z} }{|Z|}, \label{alpZ}
\ee
Then the BPS bound for the total energy is
\be
E \geq |Z|.
\ee
As has been shown by Seiberg and Witten in \cite{SeiWitt} , the vacuum expectation values $a(u)$ and $a_D(u)$ depend on the gauge-invariant complex parameter $u$, which labels the different vacua
\begin{align}
a(u)= \frac4{\p q} \celE{q}, && a_D(u)= -i \frac4{\p q} \left[ \celE{q'}-\celK{q'} \right],  \label{aaD}
\end{align}
where $q^2=\frac2{1+u}$, $q'^2=1-q^2$ and 
where $\celE{q}$ and $\celK{q}$ are complete elliptic integrals of the first and second kind. 
The complexified coupling constant in this case is
\be
\tau=\ddp{^2}{a^2}\cl{F}=\ddp{a_D}{a}=i \frac{\celK{q'}}{\celK{q}}, \label{tau}
\ee
This description of the low energy dynamics is formally valid 
in the complex $u$-plane outside a region given by the curve of marginal stability
$\im \frac{a_D(u)}{a(u)} =0$ where the spectrum of the theory changes: particles become unstable or may disappear from the spectrum completely. In fact, the effective
description should only be trusted in the region where the degrees of freedom integrated out
are heavy compared to the degrees of freedom kept in the effective action. This condition defines a
much larger region where the effective description can be trusted.
\\
In order to find a numerical solution to the BPS equations it is necessary to make some simplifying assumptions. 
The 't Hooft-Polyakov monopole \cite{tHooft,Polyakov} can be found when we impose symmetry under the diagonal $SO(3)$ subgroup of the product of rotations and global gauge transformations $SO(3) \times SO(3)_G$~
\footnote{ $SO(3)_G$ is the homomorphic image of the gauge group $SU(2)$}, 
ie. invariance under generators $\vec{K}=\vec{J}+\vec{T}$ where $\vec{J}$ generates rotations and $\vec{T}$ gauge transformations. By imposing further a $\mathbb{Z}_2$ symmetry which consists of parity plus a sign change of $\phi$ we are left with the ansatz 
\be
\phi^A=\ue^A \phi(r), \qquad
A_i^A=\epsilon^A_{ij} \ue^j \left(\frac{1-L(r)}{r} \right), \qquad
A_0^A=\ue^A b(r),
\ee
where $r=\sqrt{x^i x^i}$ is the usual distance from the origin and $\ue^A=x^A/r$ is a unit radial vector.\\ 
The electric and magnetic fields are then
\be
B_i^A= \ue _i \ue^A \frac{L^2-1}{r^2}+\cl{P}_i^A \frac{L_r}{r}, \qquad
E_i^A= -\ue_i \ue^A b_r - \cl{P}_i^A \frac{bL}{r},
\ee
with the projector $\cl{P}_i^A=\delta_i^A-\ue_i \ue^A$. The components proportional to $\ue^A \ue_i$ ($\cl{P}_i^A$) we will call abelian (nonabelian), since the abelian part is related to the abelian $U(1)$ symmetry which survives symmetry breaking.\\

Inserting in the BPS equations \eqref{BPS} one finds
\be \label{fifir}
\sqrt{2} \ue^{i \alpha} \phi_r = \frac{1-L^2}{r^2} +i b_r ,
\qquad
\sqrt{2} \ue ^{i \alpha} \phi = - \dd{}{r}\ln L +i b.
\ee
From these we can obtain the differential equation
\be
\dd{^2}{r^2} (\ln L) =\frac{L^2-1}{r^2}, 
\ee
which has the solution 
\be
L=\frac{\kappa r}{\sinh[ \kappa (r+\delta)]},
\ee
with constants $\delta$ and $\kappa$.
The constant $\kappa$ is given by considering the $r \rightarrow \infty $ limit: Taking the limit of the real part of the second BPS equation \eqref{BPS} one sees that
\be
\sqrt{2} \re \left\{ i \frac{n_m \overline{a_D} +n_e \overline{a}}{|n_m a_D+n_e a|} a\right\} = - \lim_{r \rightarrow \infty} \dd{}{r} \ln L;
\ee
the right hand side of this is 
$$\kappa = \lim_{r \rightarrow \infty}-\frac1r+\kappa \coth [\kappa (r+\delta)]. $$
Altogether $\kappa$ depends only on $u_0 = \lim_{r\rightarrow \infty}$
(for given quantum numbers) %\marginpar{dependence??}
\be
\kappa=\sqrt{2} \frac{n_m a \bar{a}}{|n_m a_D+n_e a|} \im \frac{a_D}{a}. 
\ee
For the classical 't Hooft-Polyakov monopole we require the potentials $A_i^A$ to be finite at $r=0$
which implies the condition $L \rightarrow 1$ for $r \rightarrow 0$. Then the parameter $\delta$ must be chosen to be zero. However we do not need to impose any such requirement on the quantum corrected monopole since $r=0$ always lies in the region where the theory becomes strongly coupled and we cannot trust the low energy description anymore. We therefore leave $\delta$ arbitrary. 
This does not effect the $r \rightarrow \infty$ behavior, so the magnetic quantum number is just as in the classical case $n_m=1$. \\

Defining 
\be\label{XXD}
X=\re \left(\ue^{i \alpha} \phi \right), \qquad
X_D=\re \left(\ue^{i \alpha} \phi_D\right), 
\ee
it was shown in \cite{Chalmers} that 
\be
n_m X_D(r)+n_e X(r)=0 \label{XDX},
\ee
ie.  the local central charge $Z=n_m \f_D +n_e \f$ has a constant phase 
\be
\re \ue^{i \a} Z(r)=0.
\label{phlocZ}
\ee
The same result was also derived from a string theory perspective in \cite{Fayyazuddin,AnsarGeo,Mikhailov}

The imaginary part of $\ada$ can be written using the central charge $Z$ and the field $X$ as
\be
\label{imada}
\im \ada (r)= \frac1{n_m |\phi|^2} |Z(r)| X(r).
\ee
Thus if $\im \ada =0$ at a critical radius $r_0$, this corresponds to two possibilities: either $|Z(r_0)|=0$ (a solution called Z-pole) or $X(r_0)=0$  (a solution called X-pole).

\section{The moduli space dependence}
In this section we shall derive the spatial dependence of the moduli. We know both the spatial dependence of the scalar field and the dependence  on the moduli. This will enable us to find a differential equation for the moduli.
As was mentioned before the prepotential $\cl{F}$ can depend only on 
$\phi^2=\S_{A=1}^3 \phi^A~\phi^A$. 
Taking the derivatives we find that the second derivative (the coupling) can be written as
\be
\cl{F}_{AB}= \frac{\phi_D}{\phi} \cl{P}_{AB}+\tau \, \ue_A \ue_B,
\ee
with $\tau = \cl{F}^{\prime\prime}$ given by \eqref{tau}.
Inserting the radial ansatz
in equation (\ref{Pi}) and using the fact that $\cl{P}_{AB}$ and $\ue_i\ue^B$ are orthogonal projection
operators we have 
\be
\Pi_{iA}=-\frac1{4\p} \re \left[ \cl{P}_{Ai} \left( \frac{L_r}{r}-i\frac{bL}{r} \right) \ada +\tau\, \ue_i\ue_A \left( \frac{L^2-1}{r^2}-ib_r\right) \right]\nn.
\ee
The functions $L,b$ are real, so we can write everything in terms of real and imaginary parts and obtain
\be
\Pi_{iA}=-\frac1{4\p} \left[ \cl{P}_{Ai} \left( \frac{L_r}{r}\re \ada +\frac{bL}{r} \im \ada \right) +\ue_A \ue_i \left( \frac{L^2-1}{r^2} \re \, \tau +b_r \im \,\tau \right) \right]. \label{Pi2}
\ee
On the other hand we have from the BPS equation \eqref{BPS} 
\bea
B_i^A &=& -\sqrt{2} \re \left\{ \ue^{i \alpha} \nabla_i \phi^A \right \}\nn , \\
\Pi_{iA} &=& -\frac{1}{4\pi} \re \left\{ \cl{F}_{AB} \left(B_i^B+iE_i^B\right)\right\}
 = \frac1{4\p} \sqrt{2} \re \left\{ \nabla_i\left( \ue^{i \alpha} \phi_A^D \right)\right\}
\nn ,
\eea
where we have used that $\cl{F}_{AB} \nabla_i \phi^B=\nabla_i \cl{F}_A = \nabla_i\phi^D_A$.  
In terms of the definitions \eqref{XXD} we may now write
\bea
B_i^A &=& -\sqrt{2} \nabla_i \left(\ue^A X \right),\\
\Pi_{iA}&=& \frac1{4\p}\sqrt{2}\nabla_i \left(\ue_A X_D \right).
\eea
Then multiplying equation \eqref{XDX} by $\sqrt{2} \ue_A$ and letting $\nabla_i$ operate on it , we get
\be
n_m \sqrt{2} \nabla_i \left(\ue^A X_D \right) +n_e \sqrt{2} \nabla_i \left( \ue ^A X \right) =0,\nn 
\ee
which gives us another relation between the conjugate momentum and the magnetic field
\be
\Pi_{iC}=\nmne \frac1{4\p} B_i^A \delta_{AC}. \label{Pi3}
\ee
Comparing this with \eqref{Pi2} (after inserting the radial ansatz) we are left with
\begin{gather}
\nmne \frac1{4\p} \left(\ue _i \ue_A \frac{L^2-1}{r^2}+\cl{P}_{iA} \frac{L_r}{r} \right) = \nn
\\ =-\frac1{4\p} \left(\cl{P}_{iA} \left( \frac{L_r}{r} \re \ada +\frac{bL}{r} \im \ada \right) +\ue_A \ue_i \left( \frac{L^2-1}{r^2} \re\,  \tau +b
_r \im \, \tau \right) \right) \nn.
\end{gather}
Splitting this in the real and imaginary part we have the following relations
\bea
\nmne \frac{L^2-1}{r^2} &=& - \left( \frac{L^2-1}{r^2} \re \, \tau +b_r \im \,\tau \right) ,  \\
\nmne \frac{L_r}{r} &=& - \left( \frac{L_r}{r} \re \ada +\frac{bL}{r} \im \ada \right).
\eea
Thus although we do not have the explicit dependence of $b$ on $r$ or $u$ we know that
\bea
b&=& -\frac{\nmne +\re \ada}{\im \ada} \frac{L_r}{L} \label{bb}, \\
b_r &=&-\frac{L^2-1}{r^2} \frac{ \nmne +\re \tau }{\im \tau} \nn.
\eea
When we insert this in the BPS equation \eqref{BPS}
we find
\begin{gather}
\sqrt{2} \, \ue ^{i \alpha} \phi_r =
\frac{1-L^2}{r^2} \frac{i}{ \im \tau} \left( \nmne +\overline{\tau} \right). \nn
\end{gather}
But since $\phi(r)=\phi(u(r))$, we can use the chain rule $\phi_r=\phi_u u_r$ and the dependence $\phi(u)$ in \eqref{aaD}, which gives
$$ \phi_u = \frac{q}{\p} \celK{q}. $$ 
Inserting everything in the BPS equation  we get 
$$
\sqrt{2}\, \ue ^{i \alpha} \sqrt{\frac2{1+u}} \frac{\celK{q}}{\p} u_r = \frac{1-L^2}{r^2} \frac{i}{\im \tau} \left(\nmne+\overline{\tau} \right),
$$
and we find the differential equation
\begin{equation}
u_r = \pp \sqrt{1+u} \frac{1-L^2}{r^2} \frac{\ue^{-i \alpha}}{\celK{q}} \frac{i}{\im \tau} \left( \nmne +\overline{\tau} \right). \label{ureq}
\end{equation}
This is a first order differential equation for the dependence of the moduli space parameter on the
distance $r$  to the center of the monopole. At $r=\infty$ the moduli space parameter $u$ will take the vacuum value $u_0$ of the theory. Moving towards the center of the monopole the parameter $u$ will
change according to the above differential equation.
The solutions have one integration constant, we will
choose it to be the parameter which labels the vacua, i.e. the value of $u$ at infinity $u(r\rightarrow~\infty)~=~u_0$.
From $u_0$ the constants $\alpha$ and $\kappa$ (which figures in the function $L(r)$) are
determined. \\

The dependence of the solution $u(r)$ on $\delta$ is hidden only in the function $L$ and can be removed by changing the parameter from $r$  to $X$ since from \eqref{fifir} 
$$
\sqrt2 \, \dd{X}{r}=\frac{1-L^2}{r^2}.
$$
This changes the differential equation to
\be
u_X = \pp \sqrt{\frac{1+u}2} \frac{\ue^{-i \alpha}}{\celK{q}} \frac{i}{\im \tau} 
\left( \nmne +\overline{\tau} \right). \label{ut}
\ee
So $\delta$ does not effect the shape of the curve $u(r)$, only its parametrization. \\

\subsection{The spatial dependence}
Since the shape of the $u(r)$ curve can be separated from its parametrization we begin
by studying the parametrization given by $X(r)$.
From \eqref{fifir} we find that the spatial dependence is given by
\be
X = \frac1{\sqrt2} \left( -\frac1r + \frac{\k}{\tanh(\kappa r + \kappa \delta)} \right).
\ee
\begin{figure}[htb!]
\centering
\resizebox{.49\textwidth}{!}{\includegraphics{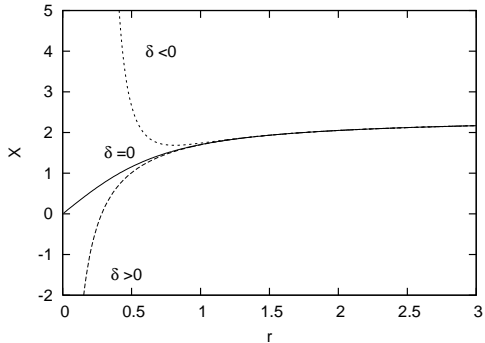}}  
\label{Xer}
\caption{The dependence X(r)}
\end{figure}
We see from fig. \ref{Xer} that there are essentially three different cases depending on whether $\delta$ is 
positive, negative or zero. 
When $\delta >0$ $X$ changes monotonically from $X=\k /\sqrt2$ at 
$r=\infty$ to $X\rightarrow -\infty$ at $r=0$. 
For $\delta=0$ $X$ also changes monotonically but
ends at $X=0$ for $r=0$. 
The $\delta < 0$ case is quite different. For $r=\infty$ it starts at $X=\k/\sqrt2$ and
decreases. For finite $r$ there is a minimum and then $X$ starts to increase and goes to $\infty$
at $r=-\delta>0$. The maximum $X$ is at an $r$ which is a solution of the condition
\be
\kappa r = \sinh [ \kappa(r+\delta)]. \label{bouncecond}
\ee
In the limit $\d \rightarrow \infty$  the $r$ dependence of $X$ is very simple $X=1/\sqrt2(\k-1/r)$. As we
will see later, in this limit, all nonabelian parts of the fields are suppressed.

\subsection{The attractor equation}
Using $X$ as a parameter and using the Seiberg-Witten metric $\ud s^2=\im \tau \ud \f \ud \bar \f$,
the equation for $u$ becomes
\be
u_X = \frac{ie^{-i\alpha}}{2n_m}g^{u\bar u}\partial_{\bar u} \bar Z
= -\frac{1}{n_m}g^{u\bar u}\partial_{\bar u}\abs{Z},
\ee
where we also have used the local central charge $Z(u) = n_m \phi_D(u) + n_e \phi(u)$ to rewrite
the equation in a suggestive form. In fact, since there is a one to one map between $u$ and $Z$
we may use $Z$ as a coordinate instead of $u$. This leads us to the equation
\be
\dd{Z}{X} = \frac{ie^{-i\alpha}}{2n_m}g^{u\bar u}\partial_u Z\partial_{\bar u}\bar Z
=\frac{ie^{-i\alpha}}{2n_m} g^{Z\bar Z},
\ee
where $g_{Z\bar Z}$ is the Seiberg-Witten metric in $Z$ coordinates, or equivalently,
using that $Z$ has constant phase for each solution, to the equation
\be
\dd{\abs{Z}}{X} = \frac{1}{2n_m}g^{u\bar u}\partial_u\abs{Z}
\partial_{\bar u}\abs{Z}.
\ee
This is an attractor equation as first discovered in \cite{SooJong,Denef1} and it can alternatively be derived
taking the zero gravity limit of the ordinary attractor equations. Using this form of the equation
and the fact that the Seiberg-Witten metric is positive definite we see that $\frac{d\abs{Z}}{dX}>0$.
This means that when $X$ is decreasing (which is the usual situation for decreasing $r$), $\abs{Z}$
will decrease and $\abs{Z}=0$ is an attractor point.

\subsection{General properties of the solutions}

Using the above relations we may write
\be
 \dd{|Z|}{r} = \frac{1}{2n_m}g^{u\bar u}\partial_u\abs{Z}
\partial_{\bar u}\abs{Z} \frac{dX}{dr},
\ee
as well as
\be
\rm{arg} \frac{dZ}{dX} =- \frac{\pi}{2}-\alpha.
\ee
{}
From \eqref{alpZ} it is clear that the phase $\alpha$ and the phase of the central charge sum up to $\pp$ 
\be
\alpha=\pp -\arg Z.
\ee
From this we can see that the curve $Z(r)$ is a straight line in the $Z$-plane going from $Z_0=Z(\infty)$ to $Z=0$. Since the derivative $\frac{dX}{dr}$ is in general positive, $\abs{Z}$ will decrease when we
decrease $r$. However, if the sign of the derivative $\frac{dX}{dr}$ changes (which is the case for 
$\delta < 0 $), the phase of the derivative jumps by $\p$ and $\abs{Z}$ starts to increase for decreasing
$r$ ending up at $X=\infty$ for $r=-\delta$. This behavior, that $\abs{Z}$ "bounces" at some value of
$r$ and starts to increase leads us to call this class of solutions, bouncing solutions.
\obrx{0.7}{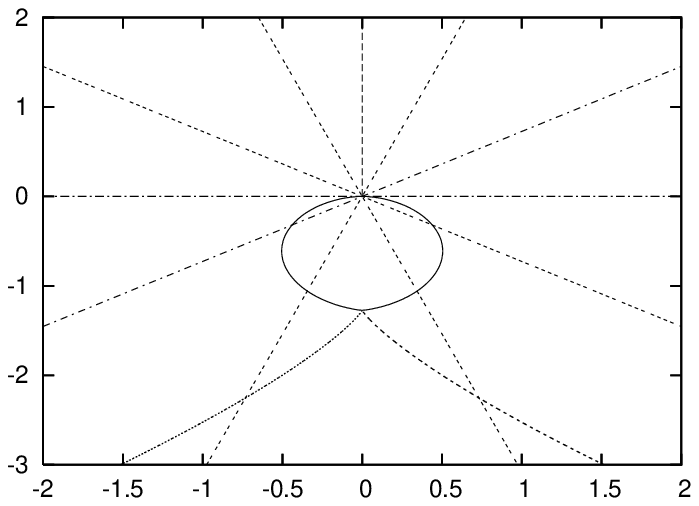}{Solutions, the curve of marginal stability and the branch cut in the complex plane of the central charge. Solutions with phase $\alpha \in (-\pp, \pp) $ hit the curve of marginal stability \Ce at $Z=0$ and are Z-poles, the others are X-poles or XZ-poles (if $\a=\pm \pp$) }{Zrov}
%
%\marginpar{Zrov.eps}
The point at which the bouncing solution turns around is given by \eqref{bouncecond}.
Whether the solution first hits $Z = 0$ or the curve of marginal stability distinguishes the X- and Z-poles.
As can be seen from fig.~\ref{Zrov} if $\alpha \in (-\pp,\pp)$ the solution is a Z-pole, if $\a= \pm \pp$ it is both an X and a Z pole, otherwise it is an X-pole. 
It follows from the explicit form of $X$ that X-pole will reach the curve of marginal stability at a radius $r_*$ given by a solution of
\be
\tanh[\kappa (r_*+\delta)] = \kappa r_*. \label{rstar}
\ee
Solutions which reach the curve of marginal stability at $r>0$ occur only for positive $\delta$. Thus we see that in the case of X-poles, the type of solution (bouncing or not) is given uniquely by the choice of the sign of $\delta$.\\

We now turn to the Z-poles. To simplify things we concentrate on the case with quantum numbers $n_m=1$, $n_e=0$ (here and for the rest of the paper). In this case the central charge is given by the dual scalar field. Also the solutions to the differential equation are symmetric with respect to the real $u$-axis, since $\left.\dd{u}{r}\right|_{u}=\left. \overline{\dd{u}{r}} \right|_{\bar{u}}$.
The dual scalar field is zero at $u=1$ so the Z-poles end at this point in the $u$-plane. We would like to see what value the parameter $r$ acquires at this point. 
At this point $\phi=4/\pi$ which corresponds to $X=4/\p \cos \a$, 
In terms of $r$ this corresponds to the value $r_{c}$, which solves
\be
-\frac1{r_c}+\kappa \coth[\kappa(r_c+\delta)]=\sqrt2 \cos\alpha \frac4{\p}. \label{rcr}
\ee
For X-poles the factor $\cos \a$ is negative and thus $X$ becomes zero before this point is reached.
For Z-poles (i.e. $\a \in (-\p/2, \p/2)$) the factor $\cos \a$ is always positive and thus $u=1$  is reached before $X=0$. 
For positive $\cos \a$ and $\delta <0$ there are in principle two possibilities: either the solution ``bounces'' back at some point or it reaches $u=1$. This is governed by the value of $\d$: for a value of $\delta$ greater than a certain $\delta_0$ the solution will reach $Z=0$ before it reaches the point of the bounce. If $\delta$ is smaller than $\delta_0$ the solution will be a bouncing solution. This particular value $\d_0<0$ solves the equation 
\be
\frac{8 \sqrt2}{\p} \frac{\kappa \cos \alpha}{\kappa^2 -\frac{32}{\p^2} \cos^2 \a}=
\sinh\left[ \frac{8\sqrt2 \cos \alpha}{\p} \frac{\kappa}{\kappa^2 -\frac{32}{\pi^2} \cos^2 \alpha } +\kappa \delta_0 \right]. \label{deltacr}
\ee

\section{Electric and magnetic fields}
As already mentioned the abelian and nonabelian parts of the electric and magnetic fields are given by
\begin{align}
B_{abel} &= -\sqrt2 \dd{X}{r}, & E_{abel}&=B_{abel} \frac{\re \t +\frac{n_e}{n_m}}{\im \t}, \label{Bfield}\\
B_{nab}&= -\sqrt2 \frac{X}r \sqrt{1-\sqrt2 r^2 \dd{X}{r}}, & E_{nab} &= B_{nab} \frac{\re \ada +\frac{n_e}{n_m}}{\im \ada}.\label{Efield}
\end{align}
In the classical case, for the 't Hooft-Polyakov monopole, the dual field is just a multiple of the scalar $\f_D=\t \f$. So, if $n_e=0$, the factors relating the abelian and the nonabelian fields are equal. Furthermore, the coupling, and thus the proportion between electric and magnetic fields, is constant. In terms of the coupling constant $g$ and the theta angle the electric fields are related to the magnetic fields as
\be
E=\frac{\theta g^2}{8 \p^2}B.
\ee

Classically, the nonabelian magnetic field is always nonzero.
In the quantum case the nonabelian magnetic field can become zero only if either $X=0$ or $\ud X/\ud r=1/(\sqrt2r^2)$. The first case is the X-pole, the other case corresponds to the $\d \rightarrow \infty$ limit. Then  $X=(\k-1/r)/\sqrt2$ and the abelian magnetic field is $B_{abel}=-1/r^2$. The abelian electric field has a more complicated dependence, due to the running coupling $\t$. Thus we can identify the $\d \rightarrow \infty$ as the abelian limit, where there are only abelian fields \cite{Denef1}.

The asymptotic behavior of the electric/magnetic fields is the same for all types of solutions. 
The abelian fields behave for large $r$ as $1/r^2$, the nonabelian fields vanish exponentially
\begin{align}
B_{abel}&\approx -\frac1{r^2}, & E_{abel}&\approx-\frac1{r^2}\frac{\re \t(u_0) +\frac{n_e}{n_m}}{\im \t(u_0)},\\
B_{nab} & \approx \frac{\k}r \ue^{-\k(r+\d)}, &
E_{nab} & \approx \frac{\k}r \ue^{-\k(r+\d)} \frac{\re \ada (u_0)+\frac{n_e}{n_m}}{\im \ada (u_0)}.
\end{align}
\section{The energy density}
The energy of a configuration is given by the Hamiltonian
\begin{equation}
H=\frac1{8\p} \im \int \ud^4x \, {\cf}_{AB} 
(E_i^A E_i^B+B_i^A B_i^B+2 \nabla_i \f^A \nabla_i \bar \f^B) ,
\end{equation}
so the energy density is
$$
{\ce}= \frac1{8 \p} \im {\cf}_{AB} 
(E_i^A E_i^B+B_i^A B_i^B+2 \nabla_i \f^A \nabla_i \bar \f^B) .
$$
For a BPS solution we see that the electromagnetic field and the Higgs field carry each one half of the total energy. We can use the BPS equation to substitute for the  Higgs field and consider
twice the electromagnetic part of the energy.
In the radial ansatz the coupling ${\cf}_{AB}$, the electric and magnetic fields split in abelian and nonabelian components. The energy density splits in an abelian and a nonabelian part as well, with  $\t$ being the abelian coupling and $ \ada$ the nonabelian coupling
\begin{align}
{\ce}=\frac1{4 \p} \im \t (B_{abel}^2 +E_{abel}^2 ) +\frac1{2 \p} \im \ada (B_{nab}^2 +E_{nab}^2),
\end{align}
the abelian and nonabelian fields are given in eq. \eqref{Efield} and \eqref{Bfield}.\\
The 't Hooft-Polyakov monopole is the classical case with ${\cf}=\half \t \f^A \f^A$ (and $\d=0$). The abelian and nonabelian coupling are the same, furthermore this coupling is fixed by its asymptotic value at infinity $\t(u_0)$.
Thus the classical energy is
\begin{align}
\ce_{cl}=\frac{\im \t}{4 \p} \left( B_{abel}^2 +E_{abel}^2 +2B_{nab}^2+2E_{nab}^2\right). 
\end{align}

The Hamiltonian can be written as the term including the BPS equation $\ch_0$ and a total derivative term, which can be rewritten as a surface term 
\begin{align*}
H &=  
H_0 -\sqrt2 \im \ue^{i \a} \int \ud^3 x \ \left(
\frac1{4 \p} \nabla_i (B^i_{A} \f_{D}^A) + \nabla_i (\P^i_{A} \f^A)\right) =\\
&= H_0 
-\sqrt2 \im \ue^{i \a} \int _{S^2_\infty} \ud^2 S_i \left( \frac1{4\p} B^i_A \f_{D}^A + \P^i_{A}\f^A\right)+\\
& +\sqrt2 \im \ue^{i \a} \int _{S^2_{r_0}} \ud^2 S_i \left( \frac1{4\p} B^i_A \f_{D}^A + \P^i_{A}\f^A\right).
\end{align*}
According to the definition of the electric and magnetic quantum numbers, the surface term at infinity is equal to $-(n_m a_D +n_e a)$. 
We can use the relation between the magnetic field and the conjugate momentum
$$
4\p n_m \P_i^A -n_e B_i^A =0,
$$
to write the third term only in terms of the magnetic field
\begin{align*}
H &= H_0 +\sqrt2 \im \ue^{i \a} (n_m a_D + n_e a)+ \sqrt2 \im \ue^{i \a} \int _{S^2_{r_0}} \ud^2 S_i B^i_A \frac1{4 \p} \left(\f_{D}^A +\frac{n_e}{n_m} \f^A \right).
\end{align*}
According to our ansatz the magnetic field splits in an abelian and a nonabelian part $B_i^A=\ue^i \ue^A (L^2-1)/r^2 +(\d_i^A-\ue^i \ue^A) L\pr/r$ and the scalar fields are pure abelian $\f^A=\f \ue^A$. Multiplying these by $\ud^2 S^i = \ud \O r^2 \ue^i$ only the abelian terms are left
\begin{align*}
H &= H_0 +\sqrt2 \im \ue^{i \a} (n_m a_D + n_e a)+ \\
&+\sqrt2 \im \ue^{i \a} \int _{S^2_{r_0}}  \frac1{4 \p n_m} \ud \O  (L^2-1) (n_m \f_D +n_e \f )= \\
&=H_0 +\sqrt2 \im \ue^{i \a} (n_m a_D + n_e a)+ \left. \sqrt2 \im \ue^{i \a} \frac1{ n_m} (L^2-1) (n_m \f_D +n_e \f )\right|_{r=r_0}.
\end{align*}
The second term includes the asymptotic value of the central charge $Z_0=Z(r=\infty)=n_m a_D + n_e a$.
The phase $\a$ was chosen in terms of this central charge as
$$
\ue^{i \a} = i \frac{\bar Z_0}{|Z_0|},
$$
so that the second term is equal to $\sqrt2 |Z_0|$ as it should be for a BPS state.
The third term represents the contribution from the inner boundary. 
Since the phase of the central charge is constant, we can rewrite the third term in much the same
way as the second term and we get for a BPS state ($H_0=0$)
\begin{align*}
H&= \sqrt2 |Z(r=\infty)|+ \sqrt2 \frac1{n_m} (L^2(r_0)-1) |Z(r_0)|.
\end{align*}
The term $L^2-1$ is up to a factor the reparametrization term $\ud X/ \ud r$. It is negative for $\d \ge 0$ with $L\rightarrow 1$ for $\d =0$ and $ r \rightarrow 0$. For negative delta, however, it can change sign: from negative (at large $r$) to positive (at small $r$).
This shows that the contribution from the inner shell lowers the total energy of the configuration. 
For Z-poles the energy of the configuration is lowered for any $r$ larger than the critical value $r_{cr}$, at which the point $u=1$ (and thus $Z=0$) is reached.
For X-poles it is lowered for all $r$ larger than the value at which the curve of marginal stability is crossed and the BPS equations do not necessarily have to hold any more.
For bouncing solutions the energy is lowered for $r$ larger than the bouncing point, however, it is increased for smaller values and tends to infinity for $r \rightarrow -\d$.

This behavior gives us a possibility to construct a completely weakly coupled monopole solution
by utilizing the properties
of the bouncing solution. If we choose $\delta$ in such a way that the value of $u$ for which
the solution turn around is in the region where we may trust the low-energy effective description,
we may cut off the solution there. The discussion above shows that there is no contribution to the
energy from the inner boundary and we may glue in a massless bubble of Higgs vacuum in the center.
That is a bubble where the Higgs field is constant radial with the value it has at the cut off point,
while the gauge field is pure gauge $A = g^{-1} d g $ with $g$ being the gauge transformation
that takes one from the Higgs field being constant and pointing in say the $z$ direction to the
radial gauge where the Higgs field point radially outwards.
Although there is no conserved charge carried by the nonabelian fields, they have nonzero
energy density. This means that there is a shell like discontinuity at the cutoff radius.

\begin{figure}[htb]
\begin{center}
\mbox{\epsfxsize=8cm\epsfbox{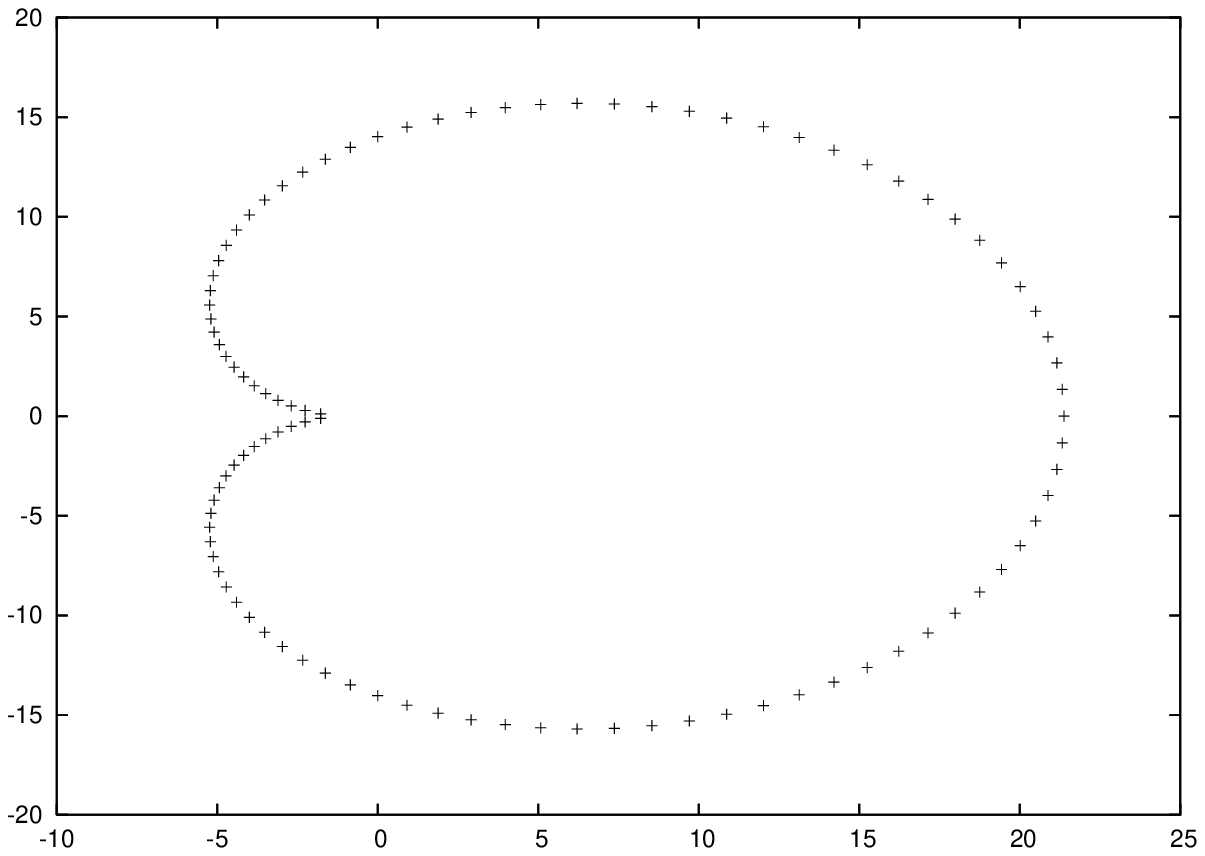}}
\end{center}
\caption{The Wilsonian core}
\label{wilson}
\end{figure}

The question of how to choose the parameter $\delta$ does not have a unique answer.
We would like to choose it so that the solution stays in the weak coupling region for all values
of $r$. Then a natural boundary is the "Wilsonian core" defined by the W bosons and the
monopoles having equal mass there. Inside the monopoles will be lighter than the W-bosons
and should thus be used as the degrees of freedom of the effective action. Outside the 
W-bosons are lighter and the standard effective action can be used. Numerically the boundary of this
region has the topology of a circle and lies outside the curve of marginal stability but touches it at one
point as in figure \ref{wilson}. There will of course be many other boundaries where other 
solitons become lighter than the W-bosons but they will not interest us here.

\section{Examples} \label{examples}
In this section we will focus on the case with quantum numbers $n_m=1$, $n_e=0$. 

\begin{figure}[htb]
\begin{center}
\mbox{\epsfxsize=6cm\epsfbox{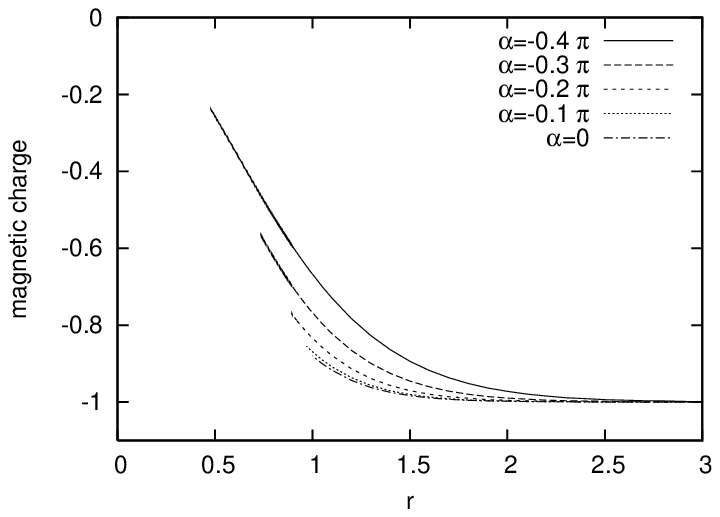}}
\mbox{\epsfxsize=6cm\epsfbox{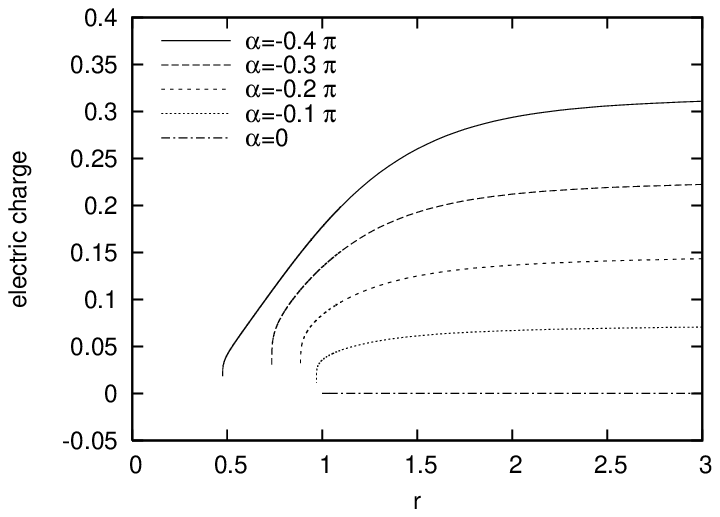}}
\end{center}
\caption{Magnetic and electric charge}
\label{fig:alphacharge}
\end{figure}

Solving the equation for the spatial dependence of the moduli numerically we may
investigate the behavior of solutions for various choices of parameters. To illustrate this,
in figure \ref{fig:alphacharge}
we give a graph which shows how the magnetic and electrical charge of the dyon changes when one
approaches the core of the monopole for various values of the phase of the central charge.
One can see in the picture that all solutions have unit magnetic charge but the electric
charge gets induced by the Witten effect and increases for solutions with increasing $|\alpha|$.

\begin{figure}[htb]
\begin{center}
\mbox{\epsfxsize=6cm\epsfbox{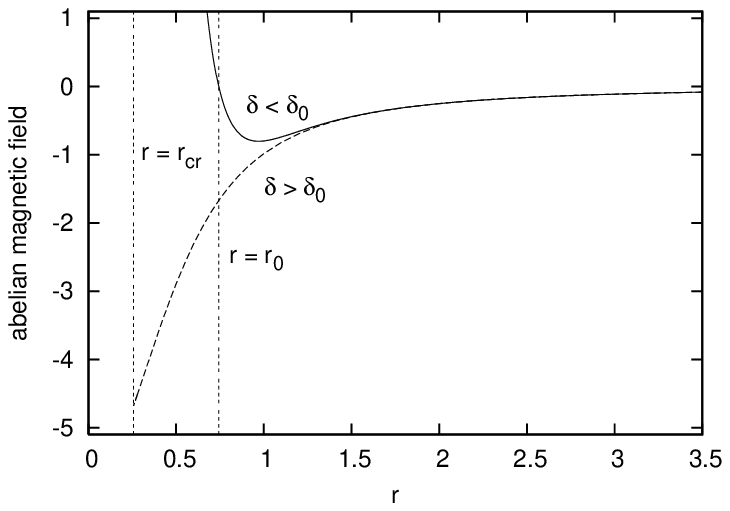}}
\mbox{\epsfxsize=6cm\epsfbox{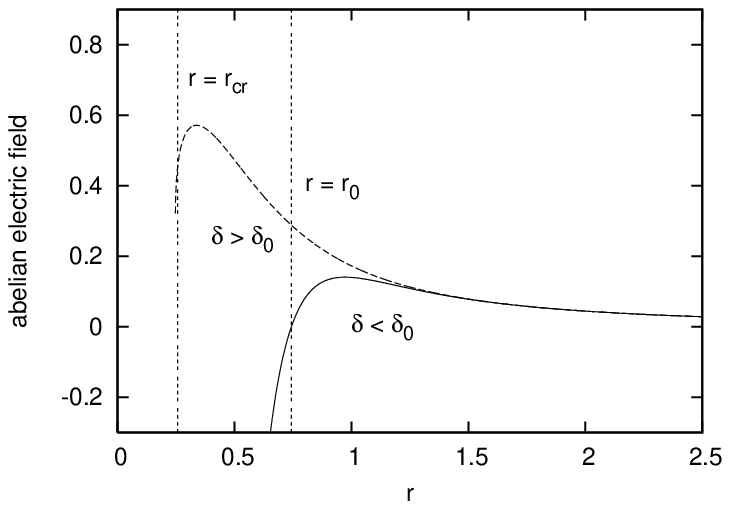}}
\end{center}
\caption{The abelian part of the magnetic and electric fields.}
\label{fig:abelian}
\end{figure}

\begin{figure}[htb]
\begin{center}
\mbox{\epsfxsize=6cm\epsfbox{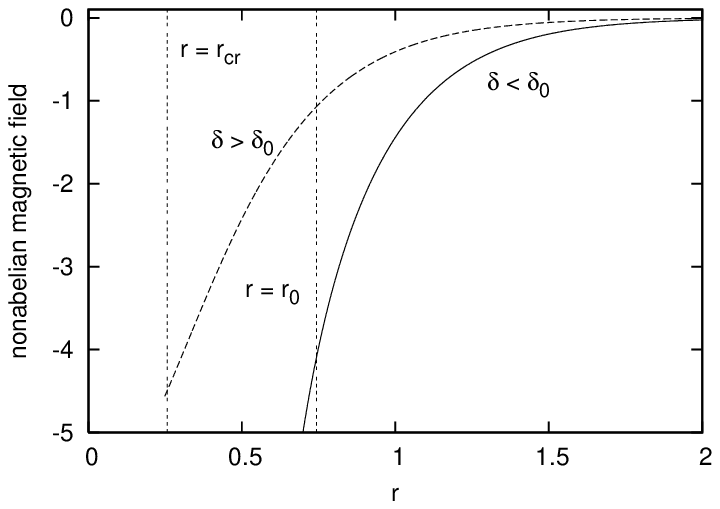}}
\mbox{\epsfxsize=6cm\epsfbox{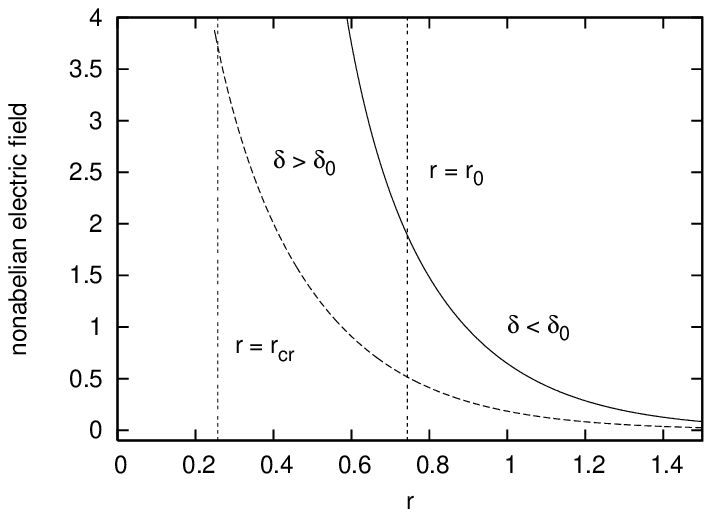}}
\end{center}
\caption{The nonabelian part of the magnetic and electric fields.}
\label{fig:nonabelian}
\end{figure}

\begin{figure}[htb]
\begin{center}
\mbox{\epsfxsize=6cm\epsfbox{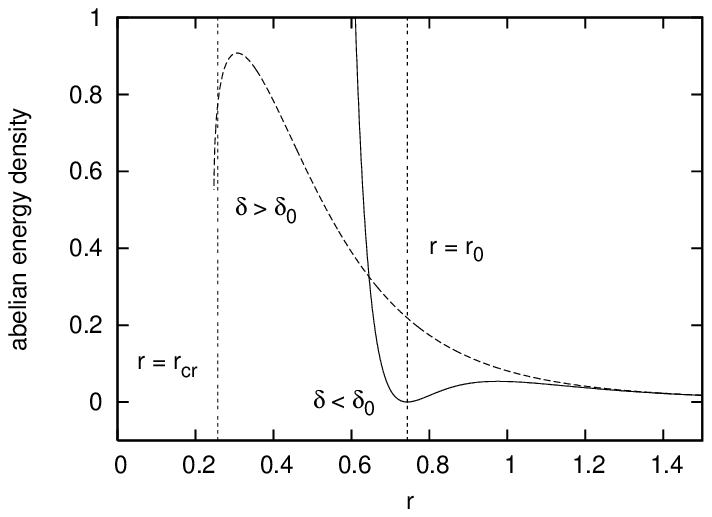}}
\mbox{\epsfxsize=6cm\epsfbox{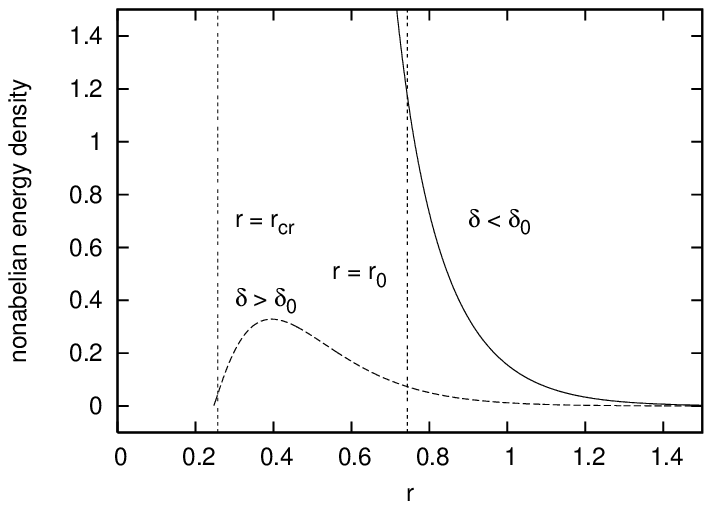}}
\end{center}
\caption{The energy density carried by the abelian and nonabelian fields.}
\label{fig:energy}
\end{figure}

We would now like to investigate the behavior of the solutions on the parameter $\delta$ and in particular
the behavior of bouncing versus non-bouncing solutions. In figures \ref{fig:abelian}, 
\ref{fig:nonabelian} and \ref{fig:energy} we therefore choose a typical Z-pole and
plot the fields for two values of $\delta$. For $\delta > \delta_0$ the solution will will hit the strong
coupling singularity at $u=1$ at finite $r$ and thus not be a bouncing
solution while for $\delta < \delta_0$ it will.

In all of the figures, the vertical line at larger $r$ represents the point of the bounce while
the inner vertical line
represents the $r$ for which the nonbouncing solution hits the strong coupling singularity
$u=1$. Notice that all the abelian fields are zero at the bouncing point
while the nonabelian fields do not seem to take any notice of the fact that the
absolute value of the moduli in fact starts to increase again.

Finally, in figure \ref{fig:energy} we display the energy density of the same two choices for
$\delta$ for the Z-pole. Notice that the energy density of the abelian components is zero
at the bouncing point while the energy density of the nonabelian fields is not.

\section{Summary and conclusions}
We have investigated properties of monopole solutions of the full low energy effective action
of $N=2$ $SU(2)$ super Yang-Mills theory. We have shown that the solutions of the quantum
corrected BPS equations are such that the local central charge always has constant phase
as one approaches the center of the monopole. We further showed that this equation can
be rewritten in a form analogous to the attractor equation and concluded that we are studying
the attractor mechanism with gravity turned off as was previously concluded in
\cite{SooJong,Denef1,Denef2}.

The general solution of the BPS equation is then such that it starts out far away from the center
at weak coupling. As one approaches the center the moduli $u$ changes toward stronger
coupling in such a way that the local central charge of the theory has constant phase. The
generic solution will at some finite radius enter the strong coupling region and may even hit
the point where the monopoles become massless. However, by choosing the integration
constants appropriately one may arrange things so that the solution is cut off while the moduli
is still in the weak coupling region and the strongly coupled center gets replaced by the weakly
coupled Higgs vacuum. Thus we have managed to find a solution to the quantum corrected BPS equations which is everywhere weakly coupled such that the effective action description can be
trusted.

The integration constant $\delta$ could in principle be considered as a new moduli for quantum BPS
monopoles. This would mean a very drastic modification to the theory of monopoles as we know
it. To get rid of this potential moduli one would need a mechanism to fix it. We have not found such
a mechanism that would uniquely fix $\delta$ in each case but we now discuss various more or less
natural choices. The first natural choice is $\delta \rightarrow \infty$  which is also considered in
\cite{Denef1}. This means that all nonabelian fields are turned off. In this scenario, since all nonabelian fields are turned off, the monopoles look more like Dirac monopoles than 't Hooft-Polyakov
monopoles.

The second natural choice is to choose $\delta$ such that the solution is a bouncing solution
that is cut off at the Wilsonian core. This is the choice we have advocated in this article. It has
the advantage that the solution lies entirely in the weakly coupled region. However, there is no
argument why we must choose exactly the Wilsonian core and not for instance a point
which lies slightly outside the Wilsonian core. Therefore, in this scenario a mechanism to fix
$\delta$ uniquely is missing.

A third natural choice of $\delta$ would be to try to choose it so that the solution hits the strong
coupling singularity at $r=0$. It is interesting to observe that this is not always possible. While
for an X-pole if we choose $\delta = 0$ we hit the curve of marginal stability (at $X=0$) when $r=0$, 
for a Z-pole, any choice of $\delta$ will give a solution that hits the strong coupling singularity
$u=1$ for $r>0$.

Another interesting question to ask is if in the gravitational attractor mechanism there exists the
equivalent of our bouncing solutions. In \cite{Denef1} it seemed that one is forced to take
$\delta\rightarrow \infty$ for the comparison with the gravitational case to work. However, this
is possibly a consequence of the fact that the gravitational side of  the problem was being purely classical.

\bigskip\bigskip
\noindent{\bf\Large Acknowledgement}:
\bigskip\bigskip

\noindent
We are grateful to S. J. Rey and M. Ro\v{c}ek for useful discussions.
The research of R.v.U. was supported by the
Czech ministry of education contract No.~MSM0021622409.

\appendix
\section{Expansions around $u=1$}
The scalar field $\f$ and its dual $\f_D$ are written in terms of elliptic integrals. The arguments of these go to respectively 0 and 1 for $u=1$. The elliptic integral $\celK{q}$ diverges for $q \rightarrow 1$, so we must use expansions.

The general formulas for expansions of the elliptic integrals around  $k=0$ and $k=1$ respectively are
\begin{gather}
\mathrm{for} \quad k\rightarrow 0  \nn\\
\celK{k}= \pp \left\{ 1+ \left(\frac{1}{2} \right)^2 k^2 +\ldots + 
\left[\frac{(2n-1)!!}{2^n n!}\right]^2 k^{2n}+\ldots \right\},  \\
\celE{k}= \pp \left\{ 1- \frac{1}{2^2} k^2 -\ldots - 
\left[\frac{(2n-1)!!}{2^n n!}\right]^2 \frac{k^{2n}}{2n-1}-\ldots \right\};\\
\mathrm{for} \quad k\rightarrow 1  \nn\\
\celK{k} = \ln \frac{4}{k '} +
\left( \frac{1}{2} \right)^2 \left( \ln \frac{4}{k '} -\frac{2}{1 \cdot 2} \right) (k') ^2+
\left( \frac{1\cdot 3}{2 \cdot 4} \right)^2 \left( \ln \frac{4}{k'}- 
\frac{2}{1 \cdot 2} - 
\right. \nn \\
 -\left. 
\frac{2}{3\cdot 4} \right) (k') ^4+ 
 \left( \frac{1\cdot 3 \cdot 5}{2 \cdot 4 \cdot 6} \right)^2 \left( \ln \frac{4}{k'}
-\frac{2}{1 \cdot 2} -\frac{2}{3\cdot 4}-\frac{2}{5\cdot 6} \right) (k') ^6+ \dots \\
\celE{k} = 1+
\frac{1}{2} \left( \ln \frac{4}{k'} -\frac{1}{1 \cdot 2} \right) (k') ^2+
\frac{1^2\cdot 3}{2^2 \cdot 4}  \left( \ln \frac{4}{k '} -\frac{2}{1 \cdot 2} -\frac{1}{3\cdot 4} \right) (k') ^4+ 
\nn \\
 + 
\frac{1^2\cdot 3^2 \cdot 5}{2^2 \cdot 4^2 \cdot 6}  \left( \ln \frac{4}{k'} 
-\frac{2}{1 \cdot 2} -\frac{2}{3\cdot 4} -\frac{1}{5 \cdot 6}\right) (k') ^6+
\ldots .
\end{gather}
We shall expand u in the form $u=1+\ve \ \eip$.
For the field $\f$ and its dual $\f_D$ we find the following expansions 
\begin{align}
\phi&=
\frac{4}{\pi} \left(1-\frac{1}{8} \ve \ln \ve \eip +\left( \frac{5}{8} \ln 2 +\frac{1}{8} \right) \ve \eip -\frac{1}{8} i \varphi \ve \eip +\frac14 \ve \eip+ O(\ve^2) \right),\\ 
\phi_D &= 
 i \half \ve \eip (1-\frac5{32} \ve \eip).  
\end{align}
We can write the expansion of $\f_D$ in the form
$
\f_D \approx\frac{i}2 \ve\eip \ue^{-\frac5{32}\ve(\cos \vf +i \sin \vf)},
$
so its phase is $\arg \f_D=\pp+\vf -\frac5{32} \ve \sin \vf$. This phase is constant along the solution of the differential equation \eqref{ureq} and equal to $\pp-\a$. Thus we get a relation between $\ve$ and $\vf$ close to $u=1$ for curves of constant $Z$-phase
\be
\a+\vf\approx \frac5{32} \ve \sin \vf.
\ee
From this we see that $\vf$ goes to $-\a$ as we get closer to $u=1$.

We can find the differential equation for $\ve$ and solve it approximately to lowest order. Inserting
$u=1+\ve \ue^{-i \a}$ in  \eqref{ut} we find the differential equation
\be
\varepsilon_t= \frac{ \frac{\pi}2 \sqrt2 } {-\frac12 \ln \varepsilon},
%\ve_t= \pp \sqrt2 \frac{\ue^{-i\a}}{-\half \ln \ve},
\ee
where we used $\celK{q} \approx -\half \ln \ve$ and $i\bar \t/\im \t\ \approx 1$.
The solution of this equation is 
\be
\ve \approx \frac{-\p\sqrt2 (t-t_0)}{\ln{[-\p\sqrt2(t-t_0)]}},
\ee
the constant $t_0$ is chosen so that $\ve(t_0)=0$, i.e. for Z-poles and critical Z-poles $t_0=t(r_c)$, for XZ-poles and bouncing XZ-poles $t_0=0$. Thus close to $u=1$ the solution goes as 
\be
u=1+\frac{-\p\sqrt2(t-t_0)}{\ln[-\p\sqrt2 ({t-t_0)]}} \ue^{-i\a}.
\ee
In terms of the parameter $r$ this can be written (except for critical Z-poles)
$$
u=1-\p \sqrt2 t_r(r_a) \frac{r-r_a}{\ln (r-r_a)} \ue^{-i \a},
$$
where $r_a$ is the point at which $u=1$. For critical Z-poles $t_r(r_a)=0$, so we have to take a higher term and get $u=1-\p \sqrt2 \frac14 t_{rr}(r_a) (r-r_a)^2/ \ln (r-r_a)$.

For the calculation of the electric fields we need the expansions of $\t$ and $\ada$, these are 
\begin{align}
\t &=-\frac{i \p}{\ln \ve +O(\ve^0)} \left(1+\frac18 \ve \eip +O(\ve^2) \right) ,\\
\ada &= \frac{i \pi}{8} \ve \eip +O(\ve^2).
\end{align}
Further we need the following expansions 
\begin{align}
\frac{\re \tau}{\im \tau} &=- \frac{-\frac18 \ve \sin \a +O(\ve^2)}{1+\frac18 \ve \cos \a+O(\ve^2)},\\
\frac{\mathcal{R}e \frac{\phi_D}{\phi}}{\mathcal{I}m \frac{\phi_D}{\phi}} &=
\tan \alpha + O(\varepsilon),\\
%\frac{\re \ada}{\im \ada} &= \frac{i\p}8\ve \ue^{-i\a}\\ 
\frac{\overline{\ada}}{\im \ada}&= -i(1+i\tan \a)+O(\ve).
\end{align}
%
%
%-----------------------------------------------------------------------
%

\end{document}